\documentstyle[twoside,fleqn,espcrc2]{article}

\newcommand{\bee} {\begin{equation}}
\newcommand{\ene} {\end{equation}}
\newcommand{\bea} {\begin{array}}
\newcommand{\ena} {\end{array}}
\newcommand{\beqa} {\begin{eqnarray}}
\newcommand{\enqa} {\end{eqnarray}}
\newcommand{\lapproxeq}{\lower .7ex\hbox{$\;\stackrel{\textstyle <}{\sim}\;$}}

\newcommand{\AmS}{{\protect\the\textfont2
  A\kern-.1667em\lower.5ex\hbox{M}\kern-.125emS}}

\title{Effective Potential Improvements,
$\epsilon$-expansions, and the Electroweak Phase
Transition}
\author{
  G. Amelino-Camelia\address{Center for Theoretical Physics, Laboratory
    of Nuclear Science and Department of Physics, \\ Massachusetts Institute
    of Technology, Cambridge, Massachusetts 02139, USA.}
       }

\begin{document}

\begin{abstract}
Two recently proposed
approaches to the study of the electroweak phase transition
are discussed.
\end{abstract}

% typeset front matter (including abstract)
\maketitle

\section{INTRODUCTION}

Recently, there has been considerable interest in the ``electroweak
phase transition" (the
temperature induced symmetry-changing phase transition\cite{linde}
of the standard electroweak model), especially in connection with the
possibility of dynamical generation of the baryon asymmetry.
An appropriate tool for the
investigation of this type of phenomena is
the finite temperature effective action\cite{jackbanf} $\Gamma_T(\phi(x))$,
where $\phi(x)$ is a trial vacuum expectation value of a
scalar\footnote{A more general definition of the effective action
can be given\cite{jackbanf}, but for the purpose of this talk it is
sufficient to consider this $\Gamma_T(\phi(x))$.}
quantum field of the theory.
$\Gamma_T(\phi(x))$
is the generating functional of the one-particle irreducible
Green's functions, and therefore encodes all physical information
about the theory, but
unfortunately, its evaluation is extremely difficult.

An important simplification can be achieved by considering
position independent trial vacuum expectation values,
$\phi(x) \!\! = \!\! constant \!\!  = \!\! \phi$.
This leads to the introduction
of the finite temperature effective potential $V_T(\phi)$,
which is related to $\Gamma_T(\phi(x))$ by
{\small
\bee
V_T(\phi) \equiv - {[\Gamma_T(\phi(x))]_{\phi=constant} \over \int dx} \,.
\label{vg}
\ene
}
\noindent $V_T(\phi)$ only encodes static\cite{jackbanf} information
about the theory (it is the generating functional of
one-particle irreducible Green's functions
at zero external momentum), but still it can be valuable in
(at least preliminary) investigations of
temperature induced phase transitions.

In spite of the simplification that follows from
considering position independent trial vacuum expectation values,
the evaluation of the effective potential $V_T(\phi)$ is still
rather difficult. It is not hard (in the imaginary time formalism
of finite temperature field theory)
to set up a loop expansion\cite{doja}
of $V_T(\phi)$ in terms of vacuum-to-vacuum
one-particle irreducible Feynman diagrams whose lines represent the
tree-level propagator,
and to evaluate (in some cases even analytically) the
first diagrams of this expansion, but near the critical temperature the
expansion is affected by infrared problems which render non-negligible
the contributions of some classes of multi-loop diagrams.
Therefore, studies of
temperature induced phase transitions cannot rely on approximations
of $V_T(\phi)$ which are obtained by truncating this ``ordinary loop
expansion" ({\it i.e.} the loop expansion discussed
in Ref.\cite{doja}).

Several alternative techniques have been proposed
for a more reliable analysis of
temperature induced symmetry-changing phase transitions.
In the following, I discuss some aspects of two of these techniques.
First, I consider improvements in the approximation scheme for the
effective potential based on selective summations of multi-loop diagrams,
and,
then, I discuss a technique for the study of
temperature induced symmetry-changing phase transitions that
uses renormalization group and $\epsilon$-expansion.
%The two approaches which have received
%more attention are based on selective summations of multi-loop diagrams
%contributing to the effective potential\cite{dinelinde}-\cite{englcjt}
%and the $\epsilon$-expansion renormalization group method.
%In the following, I shall briefly review the merits and the limits
%of these two techniques.

\section{SELECTIVE SUMMATION OF MULTI-LOOP DIAGRAMS}

Several improved approximations of the effective potential
based on selective summations of multi-loop diagrams
have been discussed in the literature (see,
for example, Refs.[4-13]);
%GAC%%%%%%%% Refs.\cite{dinelinde}-\cite{englcjt});
in particular, a way to perform systematically such
summations[8-13]
%GAC%%%%%%%% \cite{pap7}-\cite{englcjt}
can be found within the formalism of the effective
potential for composite operators\cite{jackbanf,corn} $V_T(\phi,G)$.

$V_T(\phi,G)$ is a generalization of $V_T(\phi)$
which, besides depending on a trial
vacuum expectation value of the one-point-function ($\phi$),
also depends on a trial
vacuum expectation value of the two-point-function ($G$), and is related
to $V_T(\phi)$ by
{\small
\bee
V_T(\phi) \equiv V_T(\phi,G_0) \, ,
\label{orcj}
\ene
}
\noindent
where $G_0$, which can be shown to be the full propagator of the theory,
is the solution of the stationary requirement
{\small
\bee
\biggl[{\delta V_T(\phi,G) \over \delta G}\biggr]_{G=G_0}
= 0 \,.
\label{stat}
\ene
}

Eq.(\ref{orcj}) implies that, by fixing $G \!\! = \!\! G_0$ in
the loop expansion
of $V_T(\phi,G)$
discussed in Ref.\cite{corn}, one can obtain
an ``improved loop expansion" of $V_T(\phi)$, in terms
of vacuum-to-vacuum
two-particle irreducible Feynman diagrams whose lines represent
the full propagator $G_0$.
It can be shown\cite{pap7,pap8,quiros,corn} that
approximations
of $V_T(\phi)$ which are obtained by truncating this improved loop
expansion correspond to the resummation of the classes of multi-loop
diagrams of the ordinary loop expansion
which give the
most important contributions near the critical temperature.
This gives\cite{pap8}
an important tool for the study of ``strongly first order"
phase transitions\footnote{Here a phase transition is defined to
be strongly first order if, at the critical temperature,
the value of $\phi$ at which the symmetry-breaking minimum occurs is much
greater than the product of the largest coupling of the theory
and the temperature.}, and therefore
renders possible the test of the most studied scenario\cite{ewb} for
``electroweak baryogenesis" (dynamical
generation of the baryon asymmetry at the electroweak phase
transition), which requires
the electroweak phase transition to be strongly first order.

Preliminary investigations\cite{gacewpt} of the
electroweak phase transition using this method appear to lead to
interesting results, most notably to the indication that
(in agreement with recent numerical analyses[17-19])
%GAC%%%%%%%%\cite{farrar}-\cite{shaponumb})
the possibility of baryogenesis at the one Higgs doublet electroweak
phase transition might be consistent with
the present experimental lower limit on the
Higgs mass.

Concerning the issue of reliability, it is important to realize that
this and the other approaches
based on selective summations of multi-loop diagrams are
not useful for the study of
second order or weakly first order phase transitions\cite{pap8}.
It should also be noted that there is no rigorous
argument to indicate that the critical temperature (which is obviously
also important
in determining the value of
the symmetry-breaking minimum at the critical temperature)
can be reliably estimated within these approaches\footnote{As discussed
in Refs.\cite{pibanf,pap8},
selective summations of multi-loop diagrams should lead to reliable
description of the effective potential for all values of $\phi$
greater than a certain ${\tilde{\phi}}$, but clearly, in determining
the temperature at which the symmetry breaking minimum is degenerate with
the symmetric ($\phi \! = \! 0$) one, it is necessary to
accurately describe the effective potential also for $\phi \! \sim \! 0$.};
however,
the stability[8-11]
%GAC%%%%%\cite{pap7}-\cite{gacbanf}
of such estimates with respect to higher order corrections
might be an indirect indication of their validity.

\section{$\epsilon$-EXPANSION}

Another technique that can be useful in the study of
temperature induced symmetry-changing phase-transitions
uses renormalization group and $\epsilon$-expansion\cite{march,arnold2}.
%(i.e. an expansion
%based on a dimensional continuation from 3 to 4-$\epsilon$
%spatial dimensions).
This technique exploits the fact that at high temperatures
the imaginary time formulation of finite temperature field theory
can be written as an effective 3-dimensional field theory\cite{jackbanf},
with leading dependence on the temperature introduced by the
renormalization group relations between the parameters of this 3-dimensional
theory and their 4-dimensional counterparts.
It is actually convenient\cite{arnold2} to consider
such an effective theory in 4-$\epsilon$ dimensions, obtain results
as an expansion in powers of $\epsilon$, and continue
back to the 3-dimensional theory only at the end, by taking
the $\epsilon \! \rightarrow \! 1$ limit.

Several observables relevant for the understanding of the
electroweak phase transition, and especially of electroweak baryogenesis,
have been computed\cite{arnold2} using this ``$\epsilon$-expansion method",
and it is believed that these results be reliable if the
electroweak phase transition is weakly first order.
In fact,
the ``$\epsilon$-expansion method"
has been very successful in the study
of the second order phase transition of the ``scalar $\lambda \Phi^4$ theory",
and it can therefore be reasonable to expect that it gives an accurate
description
of second order and weakly first order phase transitions.
However, the $\epsilon$-expansion is actually asymptotic\cite{arnold2}
(the terms of the expansion start growing in magnitude at orders
$n \sim 1/ \epsilon$), and this renders somewhat doubtful the meaning
of the results finally obtained taking the $\epsilon \rightarrow 1$ limit.
In some cases the $\epsilon$-expansion is useless, because already
the first terms of the expansion grow in magnitude.
Even if one finds that, in a specific calculation,
the first few terms of the expansion
do behave perturbatively,
this would still not prove (although it would be consistent with the fact)
that the results of the ``$\epsilon$-expansion method"
for that calculation are reliable.
This might be reason of concern
especially for the case of first order
phase transitions, for which (unlike the second order phase transition case)
the accuracy of the
``$\epsilon$-expansion method"
has not been satisfactorily tested.

\section{CONCLUSIONS}

A lot remains to be understood about the electroweak phase transition.
In this talk, I discussed some features and the expected (expectations
which, however, rely on arguments that still need
further investigation) reliability of two techniques
which could be useful in the study of this phase transition.
The most important observation is that
it appears that these two techniques are somewhat complementary,
one expected to be useful
in the study of strongly
first order phase transitions,
and the other expected to be useful
in the study of second order and weakly
first order phase transitions.
However,
the results of some preliminary
investigations[4,16-19,21]
%GAC%%%%%%\cite{dinelinde,gacewpt}-\cite{shaponumb,arnold2}
can be interpreted as indicating that
the (first order) electroweak phase transition might be neither strong
enough to use confidently the method of selective summations of multi-loop
diagrams contributing to the effective potential nor
weak enough to use confidently the ``$\epsilon$-expansion method".
This fact should motivate
additional and more
rigorous analysis of the limits of validity of these techniques.

\section*{ACKNOWLEDGEMENTS}

This work is supported
in part by funds provided by the U.S. Department of Energy (D.O.E.)
under cooperative agreement \#DE-FC02-94ER40818,
and by Istituto Nazionale di
Fisica Nucleare (INFN, Frascati, Italy).
I would like to thank
J. March-Russell for a very stimulating conversation on
the ``$\epsilon$-expansion method",
and
G.R. Farrar
for an informative conversation on the numerical results
presented in Refs.\cite{shaponum,shaponumb}.

\end{document}